\documentclass[11pt]{article}
\usepackage{amsmath,amssymb,color,graphics,epsfig,cite}

\usepackage{amsfonts}

\newcommand{\hoch}[1]{$\, ^{#1}$}

\newcommand{\be}{\begin{equation}}
\newcommand{\ee}{\end{equation}}
\newcommand{\bea}{\setlength\arraycolsep{2pt} \begin{eqnarray}}
\newcommand{\eea}{\end{eqnarray}}
\newcommand{\nn}{\nonumber}

\def\ft#1#2{{\textstyle{\frac{\scriptstyle #1}{\scriptstyle #2} } }}
\def\fft#1#2{{\frac{#1}{#2}}}

\def\0{{\sst{(0)}}}
\def\1{{\sst{(1)}}}
\def\2{{\sst{(2)}}}
\def\3{{\sst{(3)}}}
\def\4{{\sst{(4)}}}
\def\5{{\sst{(5)}}}
\def\6{{\sst{(6)}}}
\def\7{{\sst{(7)}}}
\def\8{{\sst{(8)}}}
\def\sst#1{{\scriptscriptstyle #1}}

\begin{document}


\begin{center}
{\Large {\bf Static and Dynamic Charged Black Holes}}

\vspace{10pt}

Hyat Huang\hoch{1}, Zhong-Ying Fan\hoch{2} and  H. L\"u\hoch{3}

\vspace{10pt}

\hoch{1} {\it Department of Physics,
Beijing Normal University, Beijing 100875, China}

\vspace{10pt}
\hoch{2}{\it Center for Astrophysics, School of Physics and Electronic Engineering, \\
 Guangzhou University, Guangzhou 510006, China }

\vspace{10pt}
\hoch{3}{\it Center for Joint Quantum Studies and Department of Physics, \\
 Tianjin University, Tianjin 300350, China}

\vspace{40pt}

\underline{ABSTRACT}
\end{center}

We consider a class of Einstein-Maxwell-dilaton theories in general dimensions and construct both static and dynamic charged black holes. We adopt the reverse engineering procedure and make a specific ansatz for the scalar field and then derive the necessary scalar potential and the non-minimal coupling function between the scalar and the Maxwell field.  The resulting static black holes contain mass and electric charge as integration constants. We find that some of the static solutions can be promoted to become dynamical ones in the Eddington-Finkelstein-like coordinates. The collapse solutions describe the evolution from a smaller charged black hole to a larger black hole state, driven by the scalar field.

\vfill {\footnotesize Corresponding author: fanzhy@gzhu.edu.cn}

\pagebreak

\tableofcontents
\addtocontents{toc}{\protect\setcounter{tocdepth}{2}}



\section{Introduction}

The proposal of holographic principle brings a new focus on the study of General Relativity in asymptotic Anti-de Sitter (AdS) geometries. According to the AdS/CFT correspondence, AdS gravity in the bulk is dual to some strongly-coupled conformal field theory (CFT) on the boundary \cite{mald,gkp,wit}. Thus AdS black holes are of great importance for the applications in the AdS/CFT. In particular, the dynamical black hole solutions are dual to non-equilibrium thermalization phenomena in quantum field theories \cite{Balasubramanian:2010ce,Balasubramanian:2011ur}.

However, the subject of dynamical solutions describing black hole collapse is hard to analyse. For example, in pure Einstein gravity, spherically-symmetric configuration cannot collapse since it involves only one conserved quantity, the mass. The breaking of spherical symmetry for collapsing can make the analysis extremely complicated and numerical methods are normally adopted. Since the scalar hair is not conserved, Einstein gravity coupled to a scalar field can have dynamical collapsing process while maintaining the spherical symmetry. In fact, even though an AdS vacua with a scalar field that satisfies the Breitenlohner-Freedman bound is stable against small perturbation at the linear level\cite{Breitenlohner:1982jf}, numerical methods indicate that it will fall into a black hole state because of the non-linear instability \cite{Bizon:2011gg,Buchel:2012uh,Wu:2013qi,Buchel:2013uba}. However, analytic dynamical black holes describing black hole formation are hard to come by. The first such example was provided by \cite{Zhang:2014sta} and further analyzed in \cite{Zhang:2014dfa}.

The construction of analytic dynamical solutions is propelled by the recent constructions of static black holes in a variety theories \cite{Henneaux:2002wm,Martinez:2004nb,Anabalon:2012dw,Gonzalez:2013aca,
Feng:2013tza,Acena:2013jya,Fan:2015tua,Fan:2015ykb,Chen:2016qks}. Some of these solutions can be promoted to become dynamical ones where the static black holes are the end points of the time evolution \cite{Fan:2015tua,Fan:2015ykb}. Further such exact dynamical solutions were later constructed in \cite{Lu:2014eta,Xu:2014xqa,Ayon-Beato:2015ada,Fan:2016yqv,Aviles:2018vnf,Xu:2019pap}.


The purpose of this paper is to construct both static and dynamic charged black holes in a class of Einstein-Maxwell-dilaton (EMD) theories in general dimensions, where the scalar field $\phi$ with a scalar potential $V(\phi)$ is non-minimally coupled to the Maxwell field, with a generic coupling function $Z(\phi)$. Our strategy is follows. We first make a simple static ansatz for the dilaton and then adopt the reverse engineering procedure to determine the metric as well as the functions $(Z,V)$. We make sure that the mass and electric charges of the solutions are integration constants rather than the specified parameters in the Lagrangian. We then rewrite these static solutions in the Eddington-Finkelstein like coordinates and construct dynamical solutions. We find that a subset of the static solutions can be promoted to become exact dynamic ones in this process.

Our paper is organized as follows. In section 2, we discuss the EMD theories in which the Maxwell field and the scalar field are coupled by a coupling function. In section 3, we construct static charged black holes in four dimensions and obtain the first law of thermodynamics. It contains spherical, toric and hypobolic topology for the solutions. In section 4, we generalize the construction to general $D$ dimensions. In section 5, we construct exact dynamical charged solutions and analyse the evolution of dynamic process. We conclude the paper in section 6.

\section{The general set up for static solutions}

In this paper, we consider a class of EMD theories in general dimensions $D$.  The Lagrangian takes the form
\be
{\cal L}=\sqrt{-g} \Big(R-\ft 12 (\partial \phi)^2-V(\phi)-\ft 14 Z(\phi)^{-1}F^2\Big)\,. \label{gen-lag}
\ee
The scalar field with potential $V(\phi)$ is minimally coupled to gravity, but non-minimally coupled to the Maxwell field with a generic function $Z(\phi)$. The Euler-Lagrange equations associated with the variation of the dilaton $\phi$, the Maxwell field $A_\mu$ and the metric $g^{\mu\nu}$ are respectively given by
\bea
&&\Box\phi =\fft{\partial V}{\partial \phi}+\fft 14 \fft{\partial Z^{-1}}{\partial \phi}F^2 \,,\cr
&&\nabla_\mu\big(\sqrt{-g}Z^{-1}F^{\mu\nu}\big) = 0\,, \cr
&&E_{\mu\nu} \equiv R_{\mu\nu}-\ft 12 R g_{\mu\nu}-T_{\mu\nu}^{A}-T_{\mu\nu}^{\phi}=0\,,
\eea
with
\bea
T_{\mu\nu}^{A} &=&\ft 12 Z^{-1}\Big(F_{\mu\nu}^2-\ft 14 g_{\mu\nu} F^2 \Big) \,,\nn\\
T_{\mu\nu}^{\phi} &=&\ft 12\partial_\mu\phi \partial_\nu \phi-\ft 12 g_{\mu\nu}\Big(\ft 12(\partial \phi)^2+V(\phi) \Big)\,.\label{TaTphi}
\eea
In this paper we first consider electrically-charged static solutions with the ansatz
\be
ds^2 = - h(r) dt^2 + \fft{dr^2}{f(r)} + r^2 d\Omega_{(D-2),k}^2\,,\qquad \phi=\phi(r)\,,\qquad
A = \xi(r) dt\,.
\ee
where $d\Omega_{(D-2),k}^2$ is a $(D-2)$-dimensional Euclidean-signatured Einstein metric with the Ricci tensor $R_{ij} = k (D-1) g_{ij}$.  The parameter $k$ can take three non-trivial discrete values, namely $k=1,0,-1$, and the corresponding maximal symmetric metrics describe the round sphere, torus and hyperboloid in $(D-2)$ dimensions.

The Maxwell equation can be solved straightforwardly, given by
\be
\xi'=\fft{Q Z}{r^{n-2}}\,\sqrt{\fft{h}{f}}\,.\label{xiprime}
\ee
where a prime is a derivative with respect to $r$.  The Einstein equations of motion are
\bea
0&=&E_{0}^0 = -\frac{(D-2)(D-3)k}{2r^{2}}+\frac{(D-2)(D-3)f}{2r^{2}}+\frac{(D-2)f'}{2r}\cr
&&\qquad\quad+\frac{Q^{2}Z}{4r^{2(D-2)}}+\frac{1}{4}f\phi'^{2}+\frac{V}{2}\,,\cr
&&\nn\\
0&=& E_{1}^1 =-\frac{(D-2)(D-3)k}{2r^{2}}+\frac{(D-2)(D-3)f}{2r^{2}}+\frac{(D-2)fh'}{2rh}\cr
&&\qquad\quad+\frac{Q^{2}Z}{4r^{2(D-2)}}-\frac{1}{4}f(r)\phi'^{2}+\frac{V}{2}\,,\cr
&&\nn\\
0&=&E_{i}^j = \Big( -\frac{(D-3)(D-4)k}{2r^{2}}+\frac{(D-3)(D-4)f}{2r^{2}}+\frac{(D-3)f'}{2r}+\frac{(D-3)fh'}{2rh}\cr
&&\qquad\quad -\frac{Q^{2}Z}{4r^{2(D-2)}}+\frac{V}{2} +\frac{f'h'}{4h}+\frac{fh''}{2h}-\frac{fh'^{2}}{4h^{2}}+\frac{1}{4}f\phi'{}^{2}\Big)\delta_{i}^j\,.
\eea
It can be easily verified that the scalar equations of motion is automatically satisfied provided that the above equations are satisfied.  A key observation is the simple expression
\be
0=E_0^0 - E_1^1 = \ft12 h \sigma^2 \Big(\phi'^2 + \fft{2(D-2)}{r} \fft{\sigma'}{\sigma}\Big)\,,\label{phisigmaeom}
\ee
where $\sigma$ is defined by
\be
f=\sigma^2 h\,.
\ee
In other words, the equation (\ref{phisigmaeom}) is independent of contributions from the Maxwell field and the scalar potential.

In this paper, we make the following ansatz for the scalar field
\begin{equation}
\phi=2k_{0}\,\text{arcsinh}\big[(\frac{q}{r})^{\Delta}\big]\,,\label{phiansatz}
\end{equation}
where $(k_0,q,\Delta)$ are constants.  This ansatz implies that the leading falloff of the scalar field at large $r$ is
\be
\phi \sim \fft{2k_0 q^\Delta}{r^\Delta}\,.\label{leadingscalar}
\ee
As we shall see later, the parameter $q$ turns out to be an integration constant and can be viewed as the scalar hair parameter or simply the scalar ``charge''. It follows from (\ref{phisigmaeom}) that
\begin{equation}
\sigma=(1+\frac{q^{2\Delta}}{r^{2\Delta}})^{\frac{k_{0}^{2}\Delta}{D-2}}.\label{sigmares}
\end{equation}
In the next section, we will reversely engineer the relevant scalar potential $V(\phi)$ and the coupling function $Z(\phi)$ and construct static black holes with dilaton $\phi$ and the metric function $\sigma$ given by (\ref{phiansatz}) and (\ref{sigmares}) respectively.

\section{Charged black holes in $D=4$ dimensions}\label{sec3}

In this section, we present the detail construction for the scalar potential $V$ and also the function $Z$, that allows us to construct black holes with (\ref{phiansatz}) and (\ref{sigmares}).  We determine the scalar potential $V$ first, by setting $Q=0$. We find that $E_{0}^0 - E^{i}_i=0$ (with no sum) implies that
\be
\ft{1}{8} r \sigma^2 \left( (r^2 \phi'^2-16) \tilde h'-4 r \tilde h''\right)-\frac{k}{r^2}=0\,,
\ee
where $\tilde h=h/r^2$.  This function can be solved up to a quadrature in terms of a hypergeometric function, namely
\be
\tilde h' = \fft{1}{r^{4}\sigma} \Big( 3\alpha - 2 k r\, {}_1F_2[- \ft{1}{2\Delta},
\ft12 \Delta k_0^2; 1-\ft{1}{2\Delta}; -\big(\ft{q}{r}\big)^{2\Delta}]\Big)\,.\label{thpeom}
\ee

\subsection{AdS planar black holes}

\subsubsection{The theories and the solutions}

The expression (\ref{thpeom}) can be integrated out easily for toroidal-symmetric solutions, corresponding to $k=0$.  In this subsection, we focus on this special case.  We find
\be
\tilde h=g^2 - \fft{\alpha q^3}{r^3} \, _2F_1[\ft{3}{2 \Delta },\ft12\Delta k_0^2;\ft{3}{2 \Delta}+1;-\left(\ft{q}{r}\right)^{2 \Delta }]\label{d4h0}
\ee
Using the remaining equations of motion, we can straightforwardly determine the scalar potential $V$, as a function of $r$.  Using (\ref{phiansatz}), we can reexpress $r$ in terms of $\phi$ and find
\bea
V(\phi) &=&-2 C^{2(\kappa-1)} (3C^2 - \kappa\Delta S^2) \Big(g^2 - \alpha S^\fft{3}{\Delta}\,
_2F_1[\ft{3}{2 \Delta },\ft12\kappa ;\ft{3}{2 \Delta}+1;-S^2]\Big)\cr
&& -6\alpha C^{\kappa} S^{\fft{3}{\Delta}}\,,
\eea
where we have defined
\be
\kappa = \Delta k_0^2\,,\qquad C \equiv \cosh\big(\ft{\phi}{2k_0}\big)\,,\qquad S\equiv\sinh\big(\ft{\phi}{2k_0}\big)\,.
\label{CSdef}
\ee
It is important to note that the parameter $q$ that appears in the metric does not appear in the potential $V$, and hence $q$ can be viewed as an integration constant of the solution.

We are now in the position to turn on the electric charge $Q$, while fixing the already obtained scalar potential $V$. Employing the analogous technique, we can write $\tilde h=\tilde{h}_0 + \tilde{h}_e$, where $\tilde{h}_0$ is given by (\ref{d4h0}).  We can solve for $\tilde{h}_e$ from $E_0^0 + E_2^2=0$ which does not involve $Z$ and $\tilde{h}_e$ is only the unknown function.  Once $\tilde{h}_e$ is determined, we can solve for $Z$ using the remaining equations of motion.  We find
\bea
Z &=& C^{2(\kappa -1)} (C^2 +\kappa\Delta S^2) \Big(\gamma_1 - \gamma_2 S^{-\fft{1}{\Delta}}
\, {}_2F_1[-\ft{1}{2\Delta}, \ft12 \kappa; 1-\ft{1}{2\Delta}; -S^2]\Big)\cr
&& +\gamma_2\, C^{\kappa} S^{-\fft{1}{\Delta}}\,,
\eea
where $\gamma_1\,,\gamma_2$ are two constants. We have thus determined the full theory, and the corresponding static solutions are given by
\bea
h &=& g^2 r^2  - \fft{\alpha q^3}{r} \, _2F_1[\ft{3}{2 \Delta },\ft12\Delta k_0^2;\ft{3}{2 \Delta}+1;-\left(\ft{q}{r}\right)^{2 \Delta }]\cr
&& +\frac{\gamma_1 Q^2}{4 r^2}-\frac{\gamma_2 Q^2}{4qr} \, _2F_1[-\ft{1}{2 \Delta },\ft12 \Delta k_0^2;1-\ft{1}{2 \Delta };-\big(\ft{q}{r}\big)^{2 \Delta }]\,.
\eea
Note that the electric potential $\xi(r)$ can be integrated out explicitly, given by
\be
\xi = -\fft{\gamma_1 Q\sigma}{r} + \fft{\gamma_2 Q}{q} \Big(\sigma\,
{}_2F_1[-\ft{1}{2\Delta}, \ft12 {\Delta k_0^2}; 1 - \ft{1}{2\Delta}; -\big(\ft{q}{r}\big)^{2\Delta}]-1\Big)\,.\label{electricpot}
\ee
Here we have chosen the gauge such that the electric potential vanishes at asymptotic infinity.

\subsubsection{Black hole thermodynamics}

The solutions constructed above all contain two integration constants, namely the scalar charge $q$ and the electric charge $Q$.  The metrics are asymptotic to the AdS spacetime in planar coordinates with the cosmological constant $\Lambda=-3g^2$.  The falloff terms of the metric function $h=-g_{tt}$ at large $r$ is
\be
h=g^2 r^2 -\fft{2M}{r} + \fft{\gamma_1 Q^2}{4r^2} + {\cal O}\Big((\ft1r)^{2\Delta+3}\Big)\,,\label{d4largerh}
\ee
where $M$ is the black holes mass, given by
\be
M=\ft12\alpha q^3 + \fft{\gamma_2 Q^2}{8 q}\,.
\ee
For simplicity, we shall consider the case with $\Delta>\ft12$ so that the last term in
(\ref{d4largerh}) falls off faster. The unusual phenomenon that $Q^2$ contributes to the total mass is related to the fact that the $\gamma_2$ terms in $Z$ diverge at the asymptotic infinity. For appropriate parameters of $(q,Q)$, the solutions describe black holes with a horizon located at the largest root $r_0$ of the function $h$, namely $h(r_0)=0$.  The temperature and entropy can be determined by the standard technique, given by
\be
T=\fft{h'(r_0) \sigma(r_0)}{4\pi}\,,\qquad S=\pi r_0^2\,.
\ee
Note that without loss of generality we assume that volume of the metric $d\Omega_{2,k=0}^2$ is $4\pi$, the same as the unit two-sphere. The electric charge and the corresponding electric potential are
\be
Q_e = \ft14 Q\,,\qquad \Phi_e = -\xi(r_0)\,.
\ee
where $\xi$ is given by (\ref{electricpot}).  In this paper, we would also like to treat the cosmological constant as the thermodynamic pressure $P$ \cite{Kastor:2009wy,Cvetic:2010jb} and the corresponding thermodynamic volume $V_{\rm th}$ can be read off from the formula obtained in \cite{Feng:2017wvc}. We have
\be
P=\fft{3g^2}{8\pi}\,,\qquad V_{\rm th} = \ft43\pi r_0^3\, \sigma(r_0)\,.
\ee
It is straightforward to verify that the first law of black hole thermodynamics is indeed satisfied
\be
dM=T dS + \Phi_e dQ_e + V_{\rm th} dP\,.\label{firstlaw}
\ee

\subsection{Spherical and general topologies}

In this subsection, we continue to study (\ref{thpeom}) with non-vanishing $k$.  For general $(\Delta, k_0)$ parameters, we do not have a closed form expression for $\tilde h$.  However, many properties can be determined even without an explicit expression.  Furthermore, for special values of $(\Delta, k_0)$, exact expressions can indeed be obtained and explicit examples will be provided.

\subsubsection{General cases}

For non-vanishing topological parameter $k$, we find
\bea
h &=& g^2 r^2 +  k \hat h \,r^2  - \fft{\alpha q^3}{r} \, _2F_1[\ft{3}{2 \Delta },\ft12\Delta k_0^2;\ft{3}{2 \Delta}+1;-\left(\ft{q}{r}\right)^{2 \Delta }]\cr
&& +\frac{\gamma_1 Q^2}{4 r^2}-\frac{\gamma_2 Q^2}{4qr} \, _2F_1[-\ft{1}{2 \Delta },\ft12 \Delta k_0^2;1-\ft{1}{2 \Delta };-\big(\ft{q}{r}\big)^{2 \Delta }]\,,
\eea
where $\hat h$ satisfies
\be
\hat h'(r)= - \fft{2}{r^3\sigma}\, _2F_1[-\ft{1}{2\Delta}, \ft12 \Delta k_0^2; 1-\ft{1}{2\Delta}; -\big(\ft{q}{r}\big)^{2\Delta}]\,.\label{hatheom}
\ee
The scalar potential is given by
\bea
V &=&-2 C^{2\Delta k_0^2-2} (3C^2 - \Delta^2 k_0^2 S^2) \Big(g^2 + \widetilde V - \alpha S^\fft{3}{\Delta}\,
_2F_1[\ft{3}{2 \Delta },\ft12\Delta k_0^2;\ft{3}{2 \Delta}+1;-S^2]\Big)\cr
&& -6\alpha C^{\Delta k_0^2} S^{\fft{3}{\Delta}}\,,
\eea
with
\be
\widetilde V = \fft{k}{q^2} \left( q^2\hat h(\phi) -\fft{C^{2-2\Delta k_0^2} S^{\fft{2}{\Delta}}}{3 C^2 - \Delta^2 k_0^2 S^2}\Big(2C^{\Delta k_0^2}\,_2F_1[-\ft{1}{2\Delta}, \ft12 \Delta k_0^2; 1-\ft{1}{2\Delta}; -S^2] +1\Big)\right)\,.
\ee
Note that since we do not know the closed expression for $\hat h(r)$ from (\ref{hatheom}), the expression of $\hat h(\phi)$ in $\widetilde V$ is unknown either.  However for special values of $\Delta$ and $k_0$, the explicit expressions can be obtained.  We shall present a few examples in the next subsection.

   The function $Z$, the relation $f=\sigma^2 h$ and electric potential $\xi$ are independent of $k$ and were given in the previous subsection.

     It is important to note that from the dimensional analysis, the $q^2 \hat h (\phi)$ is independent of $q$, which implies that $\widetilde V $ has an overall factor $k/q^2$.  Fixing $k=1$ or $k=-1$, then the parameter $q$ appears in the scalar potential and hence cannot be viewed as an integration constant of the solution any longer. This implies, in particular, that the mass, temperature and entropy are fixed for fixed charged $Q$ and the cosmological constant.  Alternatively, we can define $\beta=k/q^2$ as a fixed constant, which then implies that the topological parameter $k=q^2\beta$ becomes a continuous variable for the varying scalar charge $q$.

\subsubsection{Explicit special examples}\label{kexample}

When $\Delta=1$ and $k_0^2$ is an odd natural number, the equation (\ref{hatheom}) can be integrated out with simple functions.  We shall present a few simple examples with $k_0^2=1,3,5,7$.  We shall present both the theories and black hole solutions explicitly.

\noindent{\bf Example 1. $\Delta=1, k_0=1$:}  In this case, the $\beta$ term in $V$ vanishes automatically, and we have
\bea\label{4theory}
V &=& -\ft12 (4g^2 + 3 \alpha \phi)(2 + \cosh\phi) + \ft92 \alpha \sinh\phi\,,\cr
Z &=& \gamma_1 \cosh\phi - \gamma_2 \sinh\phi\,.
\eea
The black hole solution is given by
\bea
h&=&g^2 r^2 + k + \ft12\alpha \Big(-3 q\sqrt{r^2 +q^2}  + 3r^2{\rm arcsinh}
\big(\fft{q}{r}\big)\Big) +
\fft{(\gamma_1 q - \gamma_2 \sqrt{r^2 + q^2}) Q^2}{4q r^2}\,,\cr
f&=& \sigma^2 h\,,\quad \sigma^2 =1 + \fft{q^2}{r^2}\,,\quad \xi =
\fft{(\gamma_2 q  - \gamma_1 \sqrt{r^2 + q^2})Q}{r^2}\,,
\quad\phi=2\,{\rm arcsinh}\big(\fft{q}{r}\big)\,.
\eea
The solution is a charged generalization of the neutral solution obtained in \cite{Zhang:2014sta}.

\noindent{\bf Example 2. $\Delta=1, k_0^2=3$:}  The theory becomes more complicated, involving $(g,\alpha, \beta)$ parameters for $V$ and $(\gamma_1,\gamma_2)$ for $Z$:
\bea
V &=&-3 \cosh ^4\big(\ft{\phi }{2 \sqrt{3}}\big) (2 g^2-\sqrt{3} \alpha  \phi )\cr
&& -\ft{3}{2} \alpha  \left(\sinh \big(\ft{\sqrt{3} \phi }{2}\big)+9 \sinh \big(\ft
{\phi }{2 \sqrt{3}}\big)\right) \cosh ^3\big(\ft{\phi }{2 \sqrt{3}}\big)\cr
&&+\beta \Big[\sinh ^2\big(\ft{\phi }{2 \sqrt{3}}\big) \left(5 \cosh \big(\ft{\phi }{\sqrt{3}}\big)+\cosh \big(\ft{2 \phi }{\sqrt{3}}\big)+6\right)\cr
&&\qquad\qquad-24 \cosh ^4\big(\ft{\phi }{2 \sqrt{3}}\big) \log \left(\cosh \big(\ft{\phi }{2 \sqrt{3}}\big)\right)
\Big]\,,\nn\\
Z&=&\Big(\gamma_1 \cosh{\big(\ft{\sqrt{3}\phi}{2}\big)}-2\gamma_2 \sinh{\big(\ft{\sqrt{3}\phi}{2}\big)} \Big)\cosh^3{\big(\ft{\phi}{2\sqrt{3}}\big)}\,.
\eea
The black hole solution involving the metric functions and the electric potential is given by
\bea
h &=&g^2 r^2 + k\Big(\fft{2r^2}{q^2} \log \big(1 + \ft{q^2}{r^2}\big) - \fft{r^2}{r^2 + q^2}\Big) + 3 \alpha  r^2 \left(\frac{q}{\sqrt{q^2+r^2}}-{\rm arcsinh} \left(\frac{q}{r}\right)\right)\cr
&& +\frac{Q^2 \left(\gamma_1  q \sqrt{q^2+r^2}-\gamma_2  \left(2 q^2+r^2\right)\right)}{4 q r^2 \sqrt{q^2+r^2}}\,,\qquad k=\beta q^2\,,\cr
f &=& \big(1 + \fft{q^2}{r^2}\big)^3 h\,,\qquad \xi=\frac{\gamma_2  q Q \left(2 q^2+3 r^2\right)}{r^4}-\frac{\gamma_1  Q \left(q^2+r^2\right)^{3/2}}{r^4}\,.
\eea

\noindent{\bf Example 3. $\Delta=1, k_0^2=5$:} Interestingly, the theory becomes simpler, given by
\bea
V &=&
2 g^2 \cosh ^8\big(\ft{\sqrt{5}}{2}\phi\big) \left(\cosh(\sqrt{5} \phi)-4\right)-\ft{1}{8} \alpha \sinh ^5\big(\sqrt{5} \phi \big)\cr
&&+\fft{\beta}{18}\sinh ^6\big(\ft{\phi}{2 \sqrt{5}}\big) \left(102 \cosh \big(\ft{\phi }{\sqrt{5}}\big)+17 \cosh \big(\ft{2 \phi }{\sqrt{5}}\big)+121\right)\,,\cr
Z &=& \gamma_1 \cosh ^8\big(\ft{\phi }{2 \sqrt{5}}\big) \left(3 \cosh \big(\ft{\phi }{\sqrt{5}}\big)-2\right)\cr
&&-\ft{1}{3} \gamma_2 \left(3 \sinh \big(\ft{\sqrt{5} \phi }{2}\big)+5 \sinh \big(\ft{3 \phi }{2 \sqrt{5}}\big)\right) \cosh ^5\big(\ft{\phi }{2 \sqrt{5}}\big)\,.
\eea
The black hole solution is
\bea
h&=& g^2 r^2 + k \Big(1 - \fft{q^4(10r^2 + 9 q^2)}{9 (r^2 + q^2)^3}\Big) -\fft{\alpha q^3 r^2}{(r^2 + q^2)^{3/2}}\cr
&&+ \Big(\fft{\gamma_1}{4r^2} - \fft{\gamma_2(3r^4+12 q^2 r^2 + 8 q^4)}{12 q r^2(r^2 + q^2)^{3/2}}\Big)Q^2\,,\qquad k=\beta q^2\,,\cr
f &=& \big(1 + \fft{q^2}{r^2}\big)^5 h\,,\qquad \xi=\frac{\gamma_2  q Q \left(8 q^4+20 q^2 r^2+15 r^4\right)}{3 r^6}-\frac{\gamma_1  Q \left(q^2+r^2\right)^{5/2}}{r^6}\,.
\eea

\noindent{\bf Example 4. $\Delta=1, k_0^2=7$:}  In this case, the theory takes the form
\bea
V &=& 2 g^2 \cosh ^{12}\big(\ft{\phi }{2 \sqrt{7}}\big) \left(2 \cosh \big(\ft{\phi }{\sqrt{7}}\big)-5\right)\cr
&&-\ft{4}{5} \alpha \sinh ^5\big(\ft{\phi }{2 \sqrt{7}}\big) \cosh ^7\big(\ft{\phi }{2 \sqrt{7}}\big) \left(2 \cosh \big(\ft{\phi }{\sqrt{7}}\big)+5\right)\cr
&& + \ft{\beta}{2400} \sinh ^6\big(\ft{\phi }{2 \sqrt{7}}\big) \Big(29393 \cosh \big(\ft{\phi }{\sqrt{7}}\big)+9730 \cosh \big(\ft{2 \phi }{\sqrt{7}}\big)\cr
&&\qquad+1807 \cosh \big(\ft{3 \phi }{\sqrt{7}}\big)+139 \cosh \big(\ft{4 \phi }{\sqrt{7}}\big)+26131\Big)\,,\cr
Z &=& -\ft{2}{5} \gamma_2  \sinh \big(\ft{\phi }{2 \sqrt{7}}\big) \cosh ^7\big(\ft{\phi }{2 \sqrt{7}}\big) \Big(16 \cosh \big(\ft{\phi }{\sqrt{7}}\big)+9 \cosh \big(\ft{2 \phi }{\sqrt{7}}\big)\cr
&&\qquad +2 \cosh \big(\ft{3 \phi }{\sqrt{7}}\big)+8\Big) + \gamma_1  \cosh ^{12}\big(\ft{\phi }{2 \sqrt{7}}\big) \left(4 \cosh \big(\ft{\phi }{\sqrt{7}}\big)-3\right)\,.
\eea
The black hole solution is given by
\bea
h &=& g^2 r^2 + k \Big(1 -\frac{q^4 \left(150 q^6+611 q^4 r^2+805 q^2 r^4+350 r^6\right)}{150 \left(q^2+r^2\right)^5}\Big) -\frac{\alpha  q^3 r^2 \left(2 q^2+5 r^2\right)}{5 \left(q^2+r^2\right)^{5/2}}\cr
&& +\frac{\gamma_1 Q^2 }{4 r^2}-\frac{\gamma_2 Q^2 \left(16 q^6+40 q^4 r^2+30 q^2 r^4+5 r^6\right)}{20 q r^2 \left(q^2+r^2\right)^{5/2}}\,,\qquad k=\beta q^2\,,\cr
f &=& \big(1 + \fft{q^2}{r^2}\big)^7 h\,,\quad
\xi = \frac{\gamma_2  q Q \left(16 q^6+56 q^4 r^2+70 q^2 r^4+35 r^6\right)}{5 r^8}-\frac{\gamma_1  Q \left(q^2+r^2\right)^{7/2}}{r^8}\,.
\eea

\subsubsection{Black hole thermodynamics}

For the above explicit examples, the first law of black hole thermodynamics can be easily verified.  In fact, we can establish the first law for generic $(\Delta, k_0)$ as well, even when the explicit expression of $\tilde h$ is unknown.  This is because it is the equation (\ref{hatheom}), which is valid also for $r=r_0$, that is needed for establishing the first law.  The thermodynamical quantities are
\bea
M &=& \ft12\alpha q^3 + \fft{\gamma_2 Q^2}{8 q}\,,\qquad T=\fft{h'(r_0) \sigma(r_0)}{4\pi}\,,\qquad S=\pi r_0^2\,,\cr
Q_e &=& \ft14 Q\,,\qquad \Phi_e = -\xi(r_0)\,,\qquad P=\fft{3g^2}{8\pi}\,,\qquad V_{\rm th} = \ft43\pi r_0^3\, \sigma(r_0)\,.
\eea
Again, we have simply used the volume formula of \cite{Feng:2017wvc} to establish the thermodynamical volume $V_{\rm th}$.  One further subtlety emerges for non-vanishing $k$ is that for fixed $\beta$, we have $k=\beta q^2$.  This topological constant varies with $q$.  We would wish to restrict our attention to the case where $k$ is fixed, which requires that $q$ is fixed. With this, the first law of thermodynamics (\ref{firstlaw}) is indeed satisfied for the $k\ne 0$ case, including the spherically-symmetric charged black holes.

It is worth pointing out that when $k=1$, setting the cosmological constant parameter $g=0$ yields charged black holes that are asymptotic to the flat spacetime.

\section{Charged (AdS) black holes in $D$ dimensions}

\subsection{The general class of theories and solutions}

The four dimensional solutions constructed in the previous section can be straightforwardly generalized to those in general dimensions. For simplicity, we shall just present the final results.  We find that function $Z(\phi)$ is fully determined, given by
\bea
Z &= & \ft{1}{2}\gamma_{2}(D-2)(D-3)C^{\frac{2\kappa}{D-2}}S^{-\fft{D-3}{\Delta}}
+ C^{\frac{-2D+4\kappa+4}{D-2}}
\left(\ft{1}{2} (D-2)(D-3) C^{2}+\kappa \Delta S^{2}\right)\cr
&&\times \Big(\gamma_1 -
\gamma_2 S^{-\fft{D-3}{\Delta}}\,_{2}F_{1}[-\ft{D-3}{2\Delta},\ft{\kappa}{D-2};
1-\ft{D-3}{2\Delta};-S^{2}]\Big)\,.
\eea
The scalar potential is given by
\bea
V &=& -\ft12 C^{\frac{-2D+4\kappa+4}{D-2}} \left(2(D-2)(D-1)C^{2}-4\kappa \Delta S^{2}\right)\cr
 && \times\left(g^2 + \widetilde V-\alpha S^{\fft{D-1}{\Delta}}\,_{2}F_{1}[\ft{D-1}{2\Delta},
\ft{\kappa}{D-2};\ft{D-1}{2\Delta}+1;-S^{2}]\right)\cr
&&
-\alpha(D-2)(D-1)C^{\frac{2\kappa}{D-2}}S^{\frac{D-1}{\Delta}}\,,\cr
\widetilde V &=&  \beta \bar V(\phi)- \fft{\beta (D-2) S^{\fft{2}{\Delta}} C^{\fft{2(D-2-2\Delta)}{D-2}}}{
(D-1)(D-2) C^2 - \kappa\Delta S^2}\cr
&&\qquad\qquad\times\Big(
2C^{\fft{2\kappa}{D-2}}\, {}_2F_1[-\ft{D-3}{2\Delta}, \ft{\kappa}{D-2};
\ft{D-3}{2\Delta}-1, -S^2] + D-3\Big)\,,\label{generalV}
\eea
where the function $\bar V (\phi)$ can be expressed as a quadrature, given by
\be
\bar V (\phi) = \int d\phi\, \fft{C^{1- \fft{2\kappa}{D-2}} S^{\fft{2}{\Delta}-1}}{\kappa}
\, _2F_1[-\ft{D-3}{2\Delta}, \ft{\kappa}{D-2};
\ft{D-3}{2\Delta}-1, -S^2]\,.
\ee
The corresponding charged AdS black holes are given by
\bea
h &= & g^{2}r^{2}+ k r^2 \hat h -\fft{\alpha q^{D-1}}{r^{D-3}}\,_{2}F_{1}[\ft{D-1}{2\Delta},\ft{\Delta k_{0}^{2}}{D-2};\ft{D+2\Delta-1}{2\Delta};-\big(\ft{q}{r}\big)^{2\Delta}]\nn\\
 && +\fft{\gamma_{1}Q^{2}}{4r^{2(D-3)}} -\fft{\gamma_{2}Q^{2}}{4(r q)^{D-3}}\,_{2}F_{1}[-\ft{D-3}{2\Delta},\ft{\Delta k_{0}^{2}}{D-2};\ft{-D+2\Delta+3}{2\Delta};-\big(\ft{q}{r}\big)^{2\Delta}]\,,\nn\\
f &=& \sigma^2 h\,,\qquad \sigma=(1+\frac{q^{2\Delta}}{r^{2\Delta}})^{\frac{k_{0}^{2}\Delta}{D-2}}\,,\qquad
\phi=2k_{0}\,\text{arcsinh}\big(\frac{q}{r}\big)^{\Delta}\,,\cr
\xi &=& \fft{(D-2)\gamma_1 Q\,\sigma}{2 r^{D-3}} + \fft{(D-2)\gamma_2 Q}{2q^{D-3}}
\Big(\sigma\, _2F_1[-\ft{D-3}{2\Delta}, \ft{\kappa}{D-2}; 1-\ft{D-3}{2\Delta};
\big(\ft{q}{r}\big)^{2\Delta}]-1\Big)\,,
\eea
where
\be
\hat h' = -\fft{2}{r^3 \sigma}\, _2F_1[-\ft{D-3}{2\Delta}, \ft{\kappa}{D-2};
\ft{D-3}{2\Delta}-1, -\big(\ft{q}{r}\big)^{2\Delta}]\,.
\ee
Again, the topological parameter $k$ in general is not an independent integration constant. It is given by
\be
k=\beta q^2\,.\label{betak}
\ee
For the charged AdS planar black holes with $k=0$, the $\widetilde V$ in $V$ and $\hat h$ in $h$ can be ignored.

\subsection{Some special examples}

In spite of that the topological parameter $k$ of the black holes in general theories is not an independent integration constant, in four dimensions, there is an example of $(\Delta, k_0)$ for which $k$ is an independent parameter, namely the first example in section \ref{kexample}. The example can be generalized to higher dimensions as well. We take
\be
\Delta=D-3\,,\qquad k_{0}=\sqrt{\frac{D-2}{2(D-3)}}\,.\label{deltak0}
\ee
We find that the $\tilde V$ in (\ref{generalV}) vanishes.  The theory becomes much simpler, given by
\bea
V&=&(D-2)\big( D-2+C(2)\big) \left(-g^2+\alpha _{2}F_{1}[\ft{1}{2},\ft{D-1}{2(D-3)},\ft{3}{2}+\ft{1}{D-3};-S^2]S^{1+\ft{2}{D-3}}\right)\cr
&&-\ft{1}{2}(D-2)(D-1)\alpha S^{\ft{2}{D-3}}S(2)\,,\cr
Z &=&\ft{1}{2}(D-2)(D-3)\big(\gamma_{1}C(2)-\gamma_{2}S(2)\big)\,,
\eea
where $C(2)=\cosh(\frac{\phi}{k_{0}})$ and $S(2)=\sinh({\frac{\phi}{k_{0}}})$.  The corresponding charged black hole solutions are given by
\bea
\phi &= & 2\sqrt{\ft{D-2}{2(D-3)}}\,\text{arcsinh}\big(\frac{q}{r}\big)^{D-3}\,,\qquad
f=\big(1+ \big(\fft{q}{r}\big)^{2(D-3)}\big) h,\cr
h &= & g^{2}r^{2}+k-\fft{\alpha q^{D-1}}{r^{D-3}}\,_{2}F_{1}\big[\ft{1}{2},\ft{D-1}{2(D-3)};
1+\ft{D-1}{2(D-3)};-\big(\fft{q}{r}\big)^{2(D-3)}\big]\cr
 & & +\frac{\gamma_{1}Q^{2}}{4r^{2(D-3)}}-\frac{\gamma_{2}Q^{2}}{4(qr)^{D-3}}\,\sqrt{1+ \big(\fft{q}{r}\big)^{2(D-3)}}\,,\cr
\xi &=&\fft{(D-2)\gamma_1 Q\,\sigma}{2 r^{D-3}} + \fft{(D-2)\gamma_2 Q}{2q^{D-3}}
\Big(\sigma\, \sqrt{1-(\ft{q}{r})^{2(D-3)}}-1\Big)\,.
\eea
Here the topological parameter $k$ is a free integration constant. These solutions generalize the $\mu=0$ examples constructed in \cite{Feng:2013tza}. In five dimensions, hypergeometric function can also reduce to simpler functions, and we find
\bea
V &= & -3(g^2 + 2\alpha)\Big(3+\cosh\big(\ft{2\phi}{\sqrt3}\big)\Big) + 24\alpha
\cosh\big(\ft{\phi}{\sqrt3}\big)\Big)\,,\cr
Z &=& 3\left(\gamma_{1}\cosh\big(\ft{2\phi}{\sqrt3}\big)-
\gamma_{2}\sinh\big(\ft{2\phi}{\sqrt3}\big)\right)\,.
\eea
The solution is given by
\bea
\phi &= & \sqrt{3}\,\text{arcsinh}\big(\frac{q^{2}}{r^{2}}\big)\,,\qquad
f =  \big(1 + \fft{q^4}{r^4}\big) h\,,\cr
h &=&g^2 r^2 +  k-2\alpha\big(\sqrt{r^{4}+q^{4}} -r^2\big)\cr
&&\qquad +\frac{Q^{2}
\left(\gamma_{1}q^{2}-\gamma_{2}\sqrt{r^{4}+q^{4}}\right)}{4q^{2}r^{4}}\,,\cr
\xi &=&\fft{3\gamma_1 Q \sqrt{r^4+q^4}}{2r^4}+\fft{3\gamma_2 Q \big(\sqrt{r^8-q^8} -r^4 \big)}{2q^2 r^4}\,.
\eea

\subsection{Thermodynamics}

Following standard technique, it is straightforward to derive various thermodynamical quantities for the above solutions. We find
\bea\label{genethermo}
M &=& \ft{(D-2)\omega}{64\pi} \big(4\alpha q^{D-1} + \gamma_2 q^{3-D} Q^2\big)\,,\cr
T &=&\fft{h'(r_0) \sigma(r_0)}{4\pi}\,,\qquad S=\ft14\omega r_0^{D-2}\,,\cr
Q_e &=& \ft1{16\pi} \omega Q\,,\qquad \Phi_e = -\xi(r_0)\,,\cr
P &=& \ft{1}{16\pi} (D-1)(D-2) g^2\,,\qquad V_{\rm th} = \ft{\omega}{D-1} r_0^{D-2} \sigma(r_0)\,,
\eea
where $\omega=\int d\Omega_{(D-2)\,,k}$ stands for the volume of the codimension-two subspace. It follows that the first law of thermodynamics (\ref{firstlaw}) is satisfied.

\section{Dynamical solutions}

\subsection{Ansatz and equations of motion}

Now we turn to construct dynamical solutions whose static limits were constructed in the previous sections.  To do so, we follow the same technique of Vaidya and rewrite the static solutions in the Eddington-Finkelstein-like coordinates, namely
\be
ds^{2} = \fft{2 dr du}{\sigma(r)} -h(r) du^{2} +r^{2}d\Omega_{D-2,k}^{2}\,,
\ee
where $u=t+\int(h\sigma^{2})^{-1}dr$ is the advanced time. We now promote all the functions $(h,f,\phi, \sigma,\xi)$ to depend on the time coordinate, leading to the dynamical ansatz
\bea
ds^{2} &=& \fft{2 dr du}{\sigma(r,u)} -h(r,u) du^{2} +r^{2}d\Omega_{D-2,k}^{2}\,,\cr
\phi &=& \phi(r,u)\,,\qquad A = \xi(r,u) du\,.
\eea
The Maxwell equation implies that $\xi$ is again related to the metric via the relation (\ref{xiprime}).  The full set of independent equations of motion are now given by
\bea
E_{r}^r=0:&& 2Vr^{2(D-2)}+Q^{2}Z+2(D-3)(D-2)kr^{2(D-3)}\cr
 && +\sigma^{2}r^{2(D-3)}\left(2(D-2)rh'+h\left(2(D-3)(D-2)-
r^{2}\phi'^{2}\right)\right)=0\,,\label{eomrr}\\
E^u_r=0: && -\frac{(D-2)\sigma'}{r\sigma}-\ft{1}{2}\phi'^{2}=0\,,\label{eomur}\\
E^r_u=0:&& \sigma\left((D-2)\dot{h}+hr\dot{\phi}\phi'\right)+2(D-2)h\dot{\sigma}+
r\dot{\phi}^{2}=0\,,\label{eomru}\\
E_i^j=0:&&2\sigma^{2}r^{2D-5}\left(\sigma'\left(2(D-3)h+rh'\right)
+r\dot{\phi}\phi'\right) +4\dot{\sigma}r^{2D-4}\sigma' \cr
&& +\sigma^{3}r^{2D-6}\left(2r\left(2(D-3)h'+rh''\right)+h\left(2(D-4)(D-3)
+r^{2}\phi'^{2}\right)\right)\cr
&&-\sigma\left(2(D-4)(D-3)kr^{2D-6}+
4r^{2D-4}\dot{\sigma}'-2Vr^{2D-4}+Q^{2}Z\right)=0\,,\label{eomij}
\eea
where a prime and a dot denote a derivative with respect to $r$ and $u$ respectively.  Note that since our ansatz is most general with respect to the isometries, up to general coordinate transformations, the scalar equations of motion will be automatically satisfied once the above equations are solved.

Our technique to solve these equations is first to promote the scalar charge $q$ in the static solution to become a function of $u$, namely $q\rightarrow a(u)$. The only assumption we make is that the form of the scalar field remains unchanged, namely
\be
\phi(r,u) = 2k_{0}\,\text{arcsinh}\big[(\frac{a(u)}{r})^{\Delta}\big]\,.\label{phidynamical}
\ee
It follows from (\ref{eomur}) that the function $\sigma$ can be straightforwardly solved, given by
\begin{equation}
\sigma=\Big(1+\frac{a(u)^{2\Delta}}{r^{2\Delta}}\Big)^{\frac{k_{0}^{2}\Delta}{D-2}}\,.\label{sigmasol}
\end{equation}
It simply reduces to (\ref{sigmares}) in the static limit.   With both $\phi$ and $\sigma$ known, we note that the two equations $E_r^r=0$ and $E_i^j=0$ involve only $h, h'$ and $h''$, with no $\dot h$ or $\ddot h$. They are two consistent equations provided that $\dot a=0$. For $\dot a\ne 0$, the equations $E_r^r=0$ and $E_i^j=0$ are not always consistent.  For vanishing topological parameter $k=0$, the inconsistency can be resolved provided that
\be
\big(\Delta - \ft12 (D-2)\big)\dot a=0\,.
\ee
For solutions with arbitrary independent $k$, which can be achieved by (\ref{deltak0}), the inconsistency can be resolved provided that
\be
(D-4) \dot a=0\,.
\ee
Thus we see that for our theories and ansatz, dynamical solutions with toroidal isometry exist in general dimensions and the dynamical solution with spherical isometry exist only in $D=4$ dimensions.

Once the $h(r,u)$ solution is obtained from $E_r^r=0$ and $E_i^j=0$ equations, we can finally substitute it into $E^r_u=0$. Remarkably, we obtain a second-order ordinary differential equation of $a$ (with respect to the time coordinate $u$), for which $a=q$ is always one of the solutions.

\subsection{$D=4$}
We first consider a simple example in $D=4$ dimensions, namely the theory specified by (\ref{4theory}). The solution reads
\bea
h&=& g^{2}r^{2}+k+\frac{Q^2(\gamma_1 a-\gamma_{2}\sqrt{a^{2}+r^{2}})}{4ar^{2}} \cr
&&-\frac{2a\dot{a}}{\sqrt{a^{2}+r^{2}}}-\ft{3}{2}\alpha \Big(a\sqrt{a^{2}+r^{2}}-r^2\text{arcsinh}\big(\frac{a}{r}\big)\Big),\\
\sigma &= & \frac{\sqrt{a^{2}+r^{2}}}{r},\quad \phi = 2\text{arcsinh}\big(\frac{a}{r}\big),\quad
\xi=
\fft{(\gamma_2 a  - \gamma_1 \sqrt{r^2 + a^2})Q}{r^2}\,,
\eea
where the function $a(u)$ satisfies
\be\label{a4eve}
\dot{a}\left(12\alpha a^{4}-\gamma_{2}Q^{2}\right)+8a^{3}\ddot{a}=0.
\ee

In fact, the solution is a charged generalization of the neutral solution obtained in \cite{Zhang:2014sta}. The properties of the solution and its dynamical evolution will be analyzed in subsequent subsections.

\subsection{General dimensions}

For $k=0$, the dynamical solutions exist in general dimensions with
\be
\Delta=\ft12(D-2)\,.
\ee
The dual theories are given by
\bea\label{dytheory}
V &=& -\ft 12 (D-2) C^{2k_0^2-2}\Big( 2(D-1)C^2-(D-2)k_0^2 S^2 \Big) \\
&&\times \Big( g^2-\alpha S^{\fft{2(D-1)}{D-2}} {}_2F_1[\ft{k_0^2}{2}\,,\ft{D-1}{D-2}\,,\ft{2D-3}{D-2}\,,-S^2]\Big) -\alpha(D-1)(D-2) C^{k_0^2}S^{\fft{2(D-1)}{D-2}}\,,\nn\\
Z &=& \ft14 (D-2)C^{2k_0^2-2}\Big(2(D-3)C^2+(D-2)k_0^2 S^2 \Big)\nn\\
&& \times \Big(\gamma_1-\gamma_2 S^{-\fft{2(D-3)}{D-2}} {}_2F_1[\ft{k_0^2}{2}\,,-\ft{D-3}{D-2}\,,\ft{1}{D-2}\,,-S^2]\Big)+\ft 12(D-2)(D-3)\gamma_2 C^{k_0^2} S^{-\fft{2(D-3)}{D-2}}\,.\nn
\eea
The dynamical solutions are 
\bea\label{hdyna}
h &=&g^{2}r^{2}-2\dot{a} \big(\fft{a}{r}\big)^{D-3}
\left(1+ (\fft{a}{r})^{D-2}\right)^{-\frac{k_{0}^{2}}{2}}-\fft{\alpha a^{D-1}}{r^{D-3}}
\,_{2}F_{1}[\ft{D-1}{D-2},\ft{k_{0}^{2}}{2};\ft{2D-3}{D-2};-(\fft{a}{r})^{D-2}]\cr
 && +\fft{\gamma_{1}Q^{2}}{4r^{2(D-3)}} -\fft{\gamma_{2}Q^{2}}{4(ar)^{D-3}}
\,_{2}F_{1}[\ft{3-D}{D-2},\ft{k_{0}^{2}}{2};\ft{1}{D-2};-(\fft{a}{r})^{D-2}]\,,\cr
\sigma &=& \Big(1+(\fft{a}{r})^{D-2}\Big)^{\frac{k_{0}^{2}}{2}}\,,\qquad
\phi=2k_{0}\,\text{arcsinh}\big[\big(\fft{a}{r})^{\ft{D-2}{2}}\big]\,,\cr
\xi &=& \fft{(D-2)\gamma_1 Q\,\sigma}{2 r^{D-3}} + \fft{(D-2)\gamma_2 Q}{2a^{D-3}}
\Big(\sigma\, _2F_1[-\ft{D-3}{D-2}, \ft{\kappa}{D-2}; 1-\ft{D-3}{D-2};
\big(\ft{a}{r}\big)^{D-2}]-1\Big)\,.
\eea
The solutions generalize the neutral dynamical solutions obtained in \cite{Fan:2015ykb}. The evolution equation is given by
\be\label{evo}
\frac{\ddot{a}}{a^{2}}+\frac{\beta\dot{a}{}^{2}}{a^{3}}+\frac{\tilde{\alpha}\dot{a}}{a}-Q^{2}\tilde{\gamma}a^{3-2D}\dot{a}=0\,,
\ee
where
\be\label{para}
\beta=\ft{1}{2}\left(-(D-2)k_{0}^{2}+2D-6\right)\,,\qquad \tilde{\alpha}=\ft{1}{2}\alpha(D-1)\,,\qquad \tilde{\gamma}=\ft{1}{8}\text{\ensuremath{\gamma_{2}}}(D-3).
\ee
Notice that $\beta<D-3$ and it should not be confused with the parameter $\beta=k/q^2$ introduced in sec.\ref{sec3}. The above equation reduces to (\ref{a4eve}) in $D=4$ dimensions with $k_0=1$. Notice that in even dimensions, the dynamical solutions \eqref{hdyna} are invariant under the transformation $a\rightarrow -a\,,r\rightarrow -r$ (the invariance of $\dot a$ can be seen from \eqref{fi} or \eqref{figene}). To simply our analysis below, we will work in the coordinate $r>0$ and focus on $a\geq 0$.
\subsection{Apparent horizon for dynamic black holes}

Since the event horizon is of great difficult to solve in dynamical spacetimes, it is useful to use the apparent horizon ( or trapping horizon ) to characterize the evolution of dynamic black holes\cite{Fan:2016yqv,Wang:2003bt}. The apparent horizon is defined by
\be
\theta=0,
\ee
 where $\theta$ represents the expansion of outgoing radial null geodesic congruences, of which the tangent vector is given by
\be
\xi^\mu \ft{\partial}{\partial x^\mu}=\ft{\partial}{\partial u}+\ft{h\sigma}{2}\ft{\partial}{\partial r}.
\ee
Note the area of the hypersurface defined by constant $u$ and $r$ is given by $A=\omega r^{D-2}$ in our coordinate. The expansion can be evaluated as
\be
\theta =\ft{\xi^\mu \nabla_\mu A }{A} =\ft{(D-2)h\sigma}{2r}.
\ee
Since $\sigma$ is positive everywhere, the location of the apparent horizon is simply determined by $h=0$.

\subsection{Evolution analysis}

It turns out that the evolution equation \eqref{evo} in general can be integrated once but the result depends on the parameter $\beta$. For example, for $\beta=-2$, one finds
\be \dot a=-\tilde{\alpha} a^2\log{\big(\fft{a}{c} \big)}-\ft{\tilde{\gamma}Q^2}{2(D-2)}\,a^{6-2D}
\,,\ee
where $c$ is an integration constant. The evolution equation of this type was first examined in \cite{Fan:2016yqv}, where dynamical charged black holes are constructed in nonminimally coupled Einstein-Scalar gravity.

In the following, we shall focus on the $\beta\neq -2$ cases. We deduce
\be\label{fi}
\dot{a}=-\ft{\tilde{\alpha}}{\beta+2}a^2-\ft{\tilde{\gamma}Q^2}{2D-6-\beta}a^{6-2D}-c'\, a^{-\beta}\,,
\ee
where $c'$ is an integration constant, related to the scalar charges in the static limit. In fact, for later convenience, we can parameterize $c'$ appropriately and rewrite the first equation as follows
\be\label{figene}
\dot{a}=-\ft{\tilde{\alpha}}{\beta+2}a^2(1-\ft{c^{\beta+2}}{a^{\beta+2}})
-\ft{\tilde{\gamma}Q^2}{2D-6-\beta}a^{6-2D}\,.
\ee
On the other hand, one can read off the Vaidya mass for the dynamical black holes from \eqref{hdyna}
\be
M=\ft{(D-2)\omega}{64\pi a^4}\big(\gamma_{2}Q^2a^{7-D}+4\alpha a^{D+3}+8a^{D+1}\dot{a}\big).
\ee
Replacing $\dot{a}$ by \eqref{fi} yields
\bea
M=\ft{(D-2)\omega}{64\pi a^{D+3}}
\Big(\ft{(D-3-\beta)\gamma_2}{2D-6-\beta}\, Q^2 a^6
-\ft{4(D-3-\beta)\alpha}{\beta+2}\, a^{2D+2}-8c'\, a^{2D-\beta}\Big).
\eea
Since we are interested in the case in which the dynamical solutions describe the formation of black holes, we shall demand that the Vaidya mass is always positive definite during the whole evolution. This will constrain the specified parameters $k_0\,,\alpha\,,\gamma_2$ in the Lagrangian as well as the initial and final scalar charges and hence affect the evolution of the solutions significantly. The remaining parameter $\gamma_1$ is much relaxed, but it is still constrained by the positivity of the gauge coupling function $Z(\phi)$. We will discuss this in detail for several examples in the following.

Before moving to explicit examples, we shall further demand that in the neutral limit the mass of static black holes is nonnegative. This leads to $\alpha\geq0$, where the equality is taken when the spacetime is pure AdS in the limit.

\subsection{Explicit examples}
In this subsection, we would like to study several examples for which the first order evolution equation (\ref{fi}) can be solved analytically.

\subsubsection{ Example 1 : $D=3$ or $\gamma_2=0$ }

As a warm-up, we first examine a simpler case:  the $D=3$ dimensional solution or $\gamma_2=0$ for higher dimensional solutions. In this case, the electric charge does not affect the evolution of the scalar function $a(u)$ so that the equation (\ref{evo}) and the dynamical evolution of $a(u)$ are the same as the neutral cases studied in \cite{Fan:2015ykb}. However, a significant difference is now the initial state of the evolution is a charged naked singularity, instead of an AdS vacua or a path-dependent singularity established in \cite{Fan:2015ykb}, as will be shown later.

The first-order evolution equation in this case simplifies to
\be\label{neufi1} \dot a=-\tilde{\alpha} a^2\log{\big(\fft{a}{q} \big)}\,,\quad \mathrm{for}\quad \beta=-2\,,\ee
or
\be\label{neufi2}\dot{a}=-\ft{\tilde{\alpha}}{\beta+2}a^2(1-\ft{q^{\beta+2}}{a^{\beta+2}}) \,,\quad \mathrm{for}\quad \beta\neq-2\,,\ee
where we have set $c\equiv q$ for later purpose. It follows that from the above equations $a=q$ is always a stable point for any given $\beta$ since $\dot a>0$ when $a<q$ and $\dot a<0$ when $a>q$. This implies that $a=q$ is the scalar charge of the final stable black hole state. Furthermore, it is interesting to note that for $D=3$ or $\gamma_2=0$, the solution (\ref{hdyna}) behaves regular in the vanishing scalar charge limit $a\rightarrow 0$. Notice that in this limit, the $\dot a$ term in the metric function $h(r,u)$ vanishes as well because of $\beta<D-3$. Therefore, the initial state for any value of $\beta$ is given by
\bea\label{initial}
&&ds^2=-h(r)dr^2+2drdu+r^2 dx^idx^i\,,\quad \phi=0\,,\nn\\
&&h(r)=g^2r^2+\ft{\gamma_1 Q^2}{4r^{2D-6}}\,,\quad A=\ft{(D-2)\gamma_1 Q}{2r^{D-3}}du\,,
\eea
which describes a charged monopole with a naked singularity at the center.

The behavior of the scalar function $a(u)$ at the initial time heavily depends on the parameter $\beta$. According to the first order equations (\ref{neufi1}) and (\ref{neufi2}), one finds that for $0<\beta<D-3$, $\dot a\rightarrow +\infty$ and for $\beta=0$, $\dot a=\ft 12\tilde{\alpha}q^2>0 $ is a positive constant in the limit $a\rightarrow 0$. In these two cases, the dynamical evolution is kicked by a strong injection rate of the scalar field. On the contrary, for $\beta<0$, $\dot a=0$ at the initial time. In this case, the evolution is much slower at early times until the nonlinear effects becomes strong enough to push the dynamical process speed up. However, it is worth emphasizing that no matter how $ a(u) $ behaves initially, the Vaidya mass vanishes as well as the size of the apparent horizon since for all the cases, the initial state is described by (\ref{initial}). This is totally different from the neutral cases, in which the initial state is either an AdS vacua if $\dot a=0$ or otherwise an AdS spacetime with a path-dependent singularity at the center \cite{Fan:2015ykb}.

To end this subsection, we present analytical solutions for the scalar function $a(u)$. It turns out that for $\beta=-1$, one has
\be\label{betam1} a(u)=\fft{q e^{\tilde{\alpha}q u}}{e^{\tilde{\alpha}q u}+1} \,.\ee
From the past infinity $u\rightarrow -\infty$ to the future infinity $u\rightarrow +\infty$, the function increases monotonically, from zero to its equilibrium value $q$. It exactly describes how the spacetime spontaneously evolves from a charged naked singularity state into a stable black hole state with scalar hair.

For $\beta=-2$, the solution can be expressed as an implicit function of exponential integral function \cite{Fan:2015ykb}
\be \mathrm{Ei}\Big(\log{(\ft qa)}\Big)=-\tilde{\alpha}q u \,.\ee
While the solution is different from (\ref{betam1}), $a(u)$ behaves the same as it qualitatively.

For generic $\beta\neq -1\,,-2$, the solutions are given by \cite{Fan:2015ykb}
 \be \tilde{\alpha}q (u-u_i)=\fft{\beta+2}{\beta+1}\Big(\fft{a}{q}\Big)^{\beta+1}
 {}_2F_1\Big[1\,,\ft{\beta+1}{\beta+2}\,,\ft{2\beta+3}{\beta+2}\,,\ft{a^{\beta+2}}{q^{\beta+2}}\Big]\,,\label{solgene}\ee
where $u_i$ is an integration constant which can be properly chosen to guarantee the reality of the solutions. A simple case is for $\beta=0$, one has $a=q\, \mathrm{tanh}\big(\ft 12\tilde\alpha q u \big)$. For more details, we refer the readers to \cite{Fan:2015ykb}.

\subsubsection{ Example 2 : $D=4\,,k_0=1$}

Now we would like to study analytical examples in which the electric charge nontrivially affects the dynamical evolution of the solutions. We consider $D=4$ dimensional solutions and take $k_0=1$, which corresponds to $\beta=0$ in four-dimensions. The Vaidya mass is given by (we will set $\omega=1$ in the following)
\bea\label{d4vaidyamass}
M
&=&-\fft{\alpha}{16\pi a}\Big( a^4-3c^2 a^2-\fft{\gamma_2 Q^2}{4\alpha} \Big)\nn\\
&\equiv&-\fft{\alpha}{16\pi a}\big( a^2-\hat{q}_+^2 \big)\big( a^2-\hat{q}_-^2 \big)\,,
\eea
where
\be \hat{q}_\pm=\fft{\sqrt{6}}{2}\sqrt{c^2\pm\sqrt{c^4+\ft{\gamma_2 Q^2}{9\alpha}}} \,.\ee
It turns out that the positivity of the Vaidya mass constrains the scalar function $a(u)$ as
\bea\label{d4scalar}
&&0<a<\hat{q}_+\,,\quad \mathrm{for}\quad \gamma_2>0 \,,\nn\\
&&\hat{q}_-<a<\hat{q}_+\,,\quad \mathrm{for}\quad \gamma_2<0\,.
\eea
Interestingly, for $\gamma_2<0$ the electric charge should be bounded above $Q^2\leq |9\alpha c^4/\gamma_2|$.

The first-order equation \eqref{figene} simplifies to
\be\label{fid4}
\dot{a}=-\frac{\tilde{\alpha}}{2}(a^2-c^2)-\fft{Q^2\tilde{\gamma}}{2a^2}\,,
\ee
which can be written more compactly as
\be\label{firsta4} \dot a=-\fft{3\alpha}{4a^2}\big(a^2-q_+^2 \big)\big(a^2-q_-^2 \big)\,,\ee
where
\be q_{\pm}=\fft {\sqrt{2}}{2}\sqrt{c^2 \pm \sqrt{c^4-\ft{\gamma_2 Q^2}{3\alpha}} } \,.\ee
It is easy to see that the number of stationary points $\dot a=0$ depends on the sign of the parameter $\gamma_2$. This greatly effects the evolution of the solutions under consideration and we will discuss the $\gamma_2>0$ case and the $\gamma_2<0$ case separately in the following.

\textbf{Case 1: $\gamma_2>0$}. It is easy to see that for $\gamma_2>0$, the existence of stationary points requires that the electric charge should be bounded above as $Q^2\leq 3\alpha c^4/\gamma_2$. In general, there are two stationary points, given by $a=q_-$ and $a=q_+$ (notice that $q_-<q_+<\hat{q}_+$). However, $q_-$ is an unstable point since $\dot{a}>0$ when $a>q_-$ whilst $q_+$ is a stable point since $\dot{a}>0$ when $a<q_+$ and $\dot a<0$ when $a>q_+$. Furthermore, one always has $\dot a\geq 0$ in the regime $q_-\leq a\leq q_+$. Moreover, from (\ref{d4vaidyamass}) one finds
\be \fft{dM}{du}=\fft{\dot{a}^2}{4\pi}\geq 0 \,.\ee
Thus, both the scalar function and the Vaidya mass monotonically increase with the advanced time $u$. Without knowing the analytical expression of the scalar function $a(u)$, we may conclude that the dynamical solution describes the spacetime evolving from a smaller charged black hole with $a=q_-$ to a bigger one with $a=q_+$.
\begin{figure}
  \centering
  \includegraphics[width=200pt]{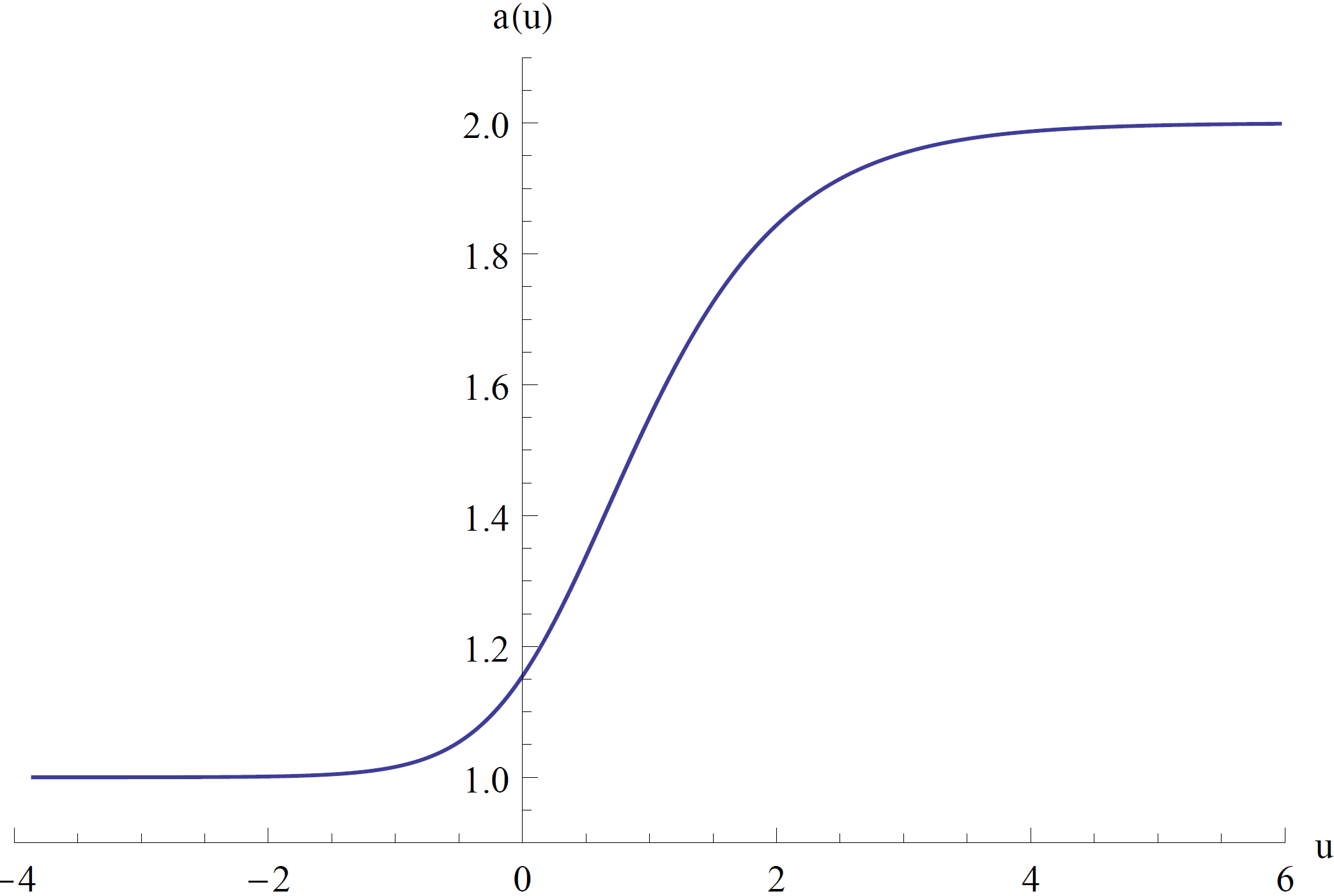}
  \includegraphics[width=200pt]{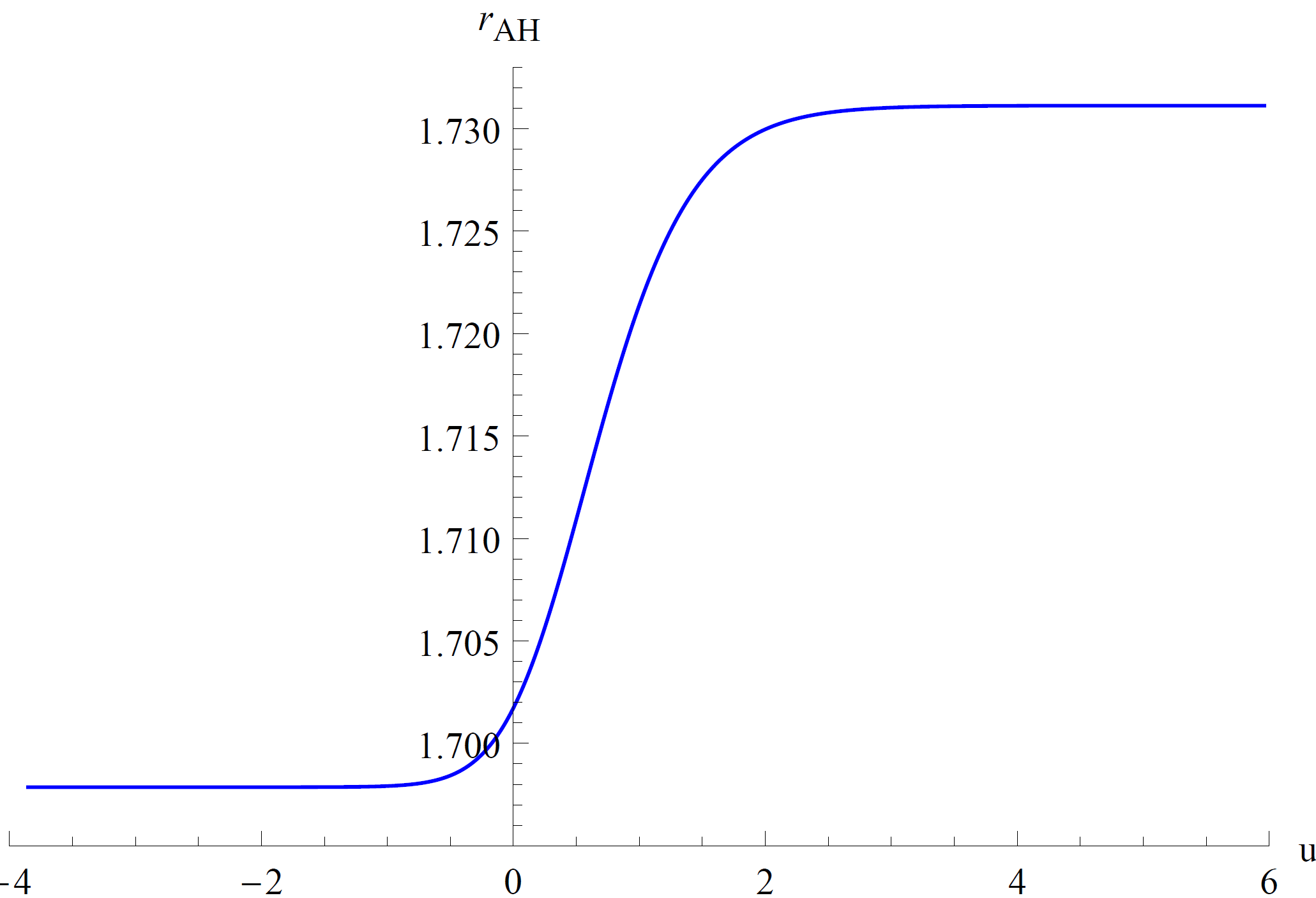}
  \caption{Left panel: The evolution of the scalar function $a(u)$ for the four dimensional solution with $k_0=1$ and $\gamma_2>0$. We have set $\gamma_2=1\,,\alpha=1/\sqrt{3}\,,c=\sqrt{5}\,,Q=4$ so that $q_-=1\,,q_+=2$. Right panel: The evolution of the apparent horizon. We further set $g=1\,,\gamma_1=9/10$. The location of the event/apparent horizon for the initial and final stable black hole are
  $r_{h_1}\simeq 1.6978\,,r_{h_2}\simeq 1.7311$, respectively.}
  \label{d4}
\end{figure}

To confirm our analysis above, we further solve the first-order equation \eqref{fid4} analytically. The solution can be written as an implicit function of $u$
\be\label{d4jie}
\ft34 \alpha\big(q_+^2-q_-^2 \big)(u-u_0)=q_+\, \mathrm{arctanh}\big(\fft{a}{q_+} \big)
-q_-\, \mathrm{arctanh}\big(\fft{q_-}{a} \big)\,,
\ee
where $u_0$ is an integration constant, which can be set to zero without loss of generality. It is immediately seen that the initial time of the evolution is at the past infinity $u\rightarrow -\infty$ whilst the final state is achieved at the future infinity $u\rightarrow +\infty$. The full evolution of the scalar function $a(u)$ is depicted as a function of the advanced time $u$ in Fig.\ref{d4}. Initially, the scalar function $a(u)$ behaves as
\be a(u)=q_-+2q_-p_-\, \mathrm{exp}\big(\ft{3\alpha(q_+^2-q_-^2)}{2q_-}\,u \big)+\cdots \,,\ee
where $p_-=\big(\fft{q_+-q_-}{q_++q_-} \big)^{q_+/q_-}$ is a positive constant. It is easily seen that $\dot a>0$ for $a>q_-$, which implies that while the initial black hole is stable against small perturbations, non-linear effects will push it to evolve into a bigger black hole. At the future infinity, the function $a(u)$ behaves as
\be a(u)=q_+-2q_+p_+\, \mathrm{exp}\big(-\ft{3\alpha(q_+^2-q_-^2)}{2q_+}\,u \big)+\cdots \,,\ee
where $p_+=\big(\fft{q_++q_-}{q_+-q_-} \big)^{q_-/q_+}$ is a positive constant. Despite that the dynamical process takes infinite times to evolve into a stable black hole state, it will speed up at late times and approach the static configuration exponentially with the relaxation time given by $\tau=2q_+/\big( 3\alpha(q_+^2-q_-^2)\big)$. In Fig.\ref{d4}, we also plot the apparent horizon as a function of the advanced time. It is immediately seen that during the whole dynamic process, it grows monotonically from the event horizon of the initial black hole to that of the final stable black hole.

In conclusion, these results together support that our dynamical solution describes how a smaller charged black hole with scalar hair undergoes nonlinear instability and spontaneously evolves into a bigger black hole state.

\textbf{Case 2: $\gamma_2<0$}. Now we move to the $\gamma_2<0$ case. In this case, $q_-$ becomes pure imaginary in the first order equation (\ref{firsta4}). As a consequence, there is only one stationary point at $a=q_+$. Again, this is a stable point corresponding to the final stable black hole. We set $q_-\equiv i \tilde{q}_-$. Then the solution (\ref{d4jie}) can be rewritten as
\be\label{d4jie2}
\ft34 \alpha\big(q_+^2+\tilde{q}_-^2 \big)(u-u_0)=q_+\, \mathrm{arctanh}\big(\fft{a}{q_+} \big)
+\tilde{q}_-\, \mathrm{arctan}\big(\fft{\tilde{q}_-}{a} \big)\,,
 \ee
where we have adopted an identity $\text{arctanh}(ix)=i \arctan(x)$.

\begin{figure}
  \centering
  \includegraphics[width=210pt]{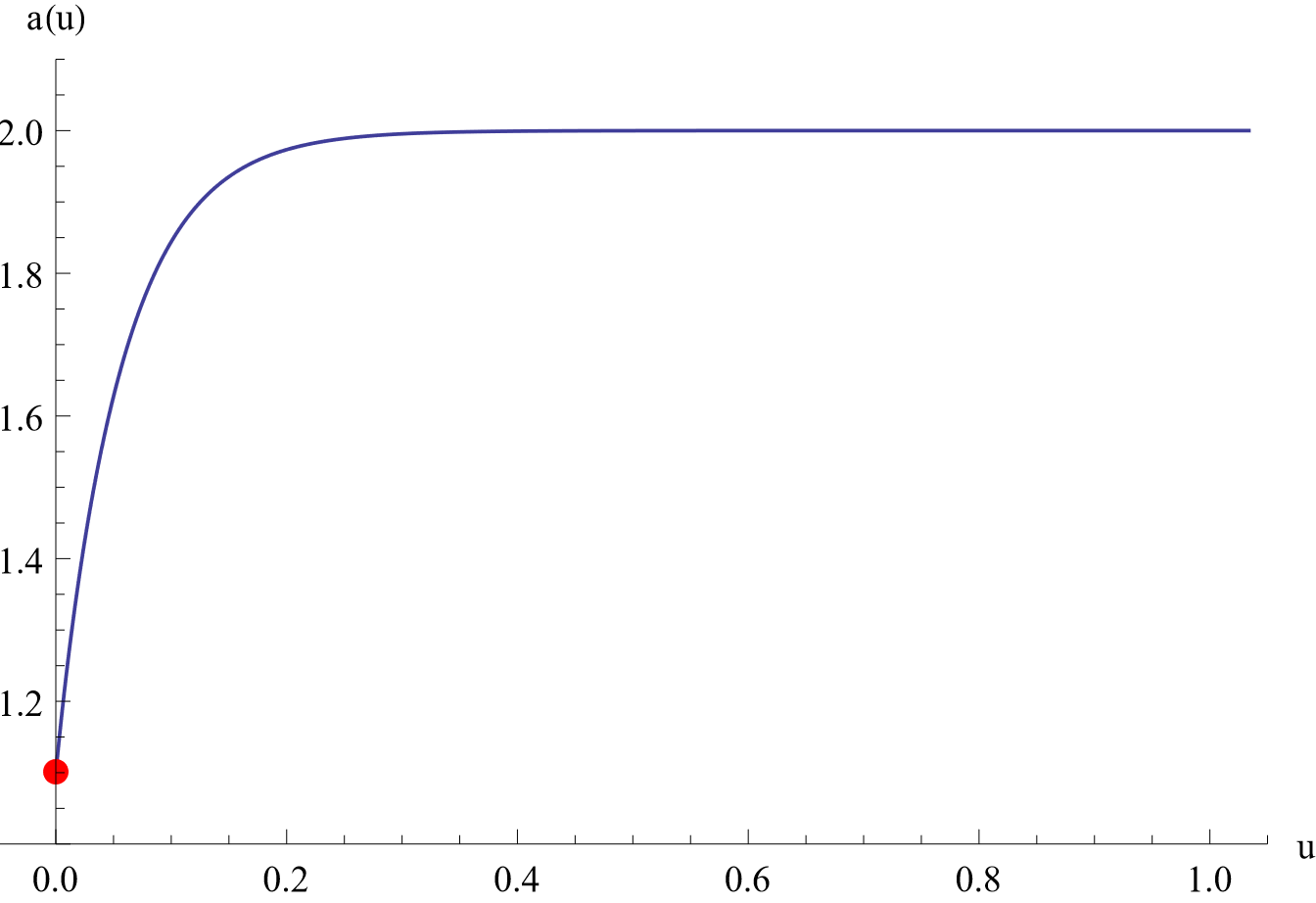}
  \includegraphics[width=210pt]{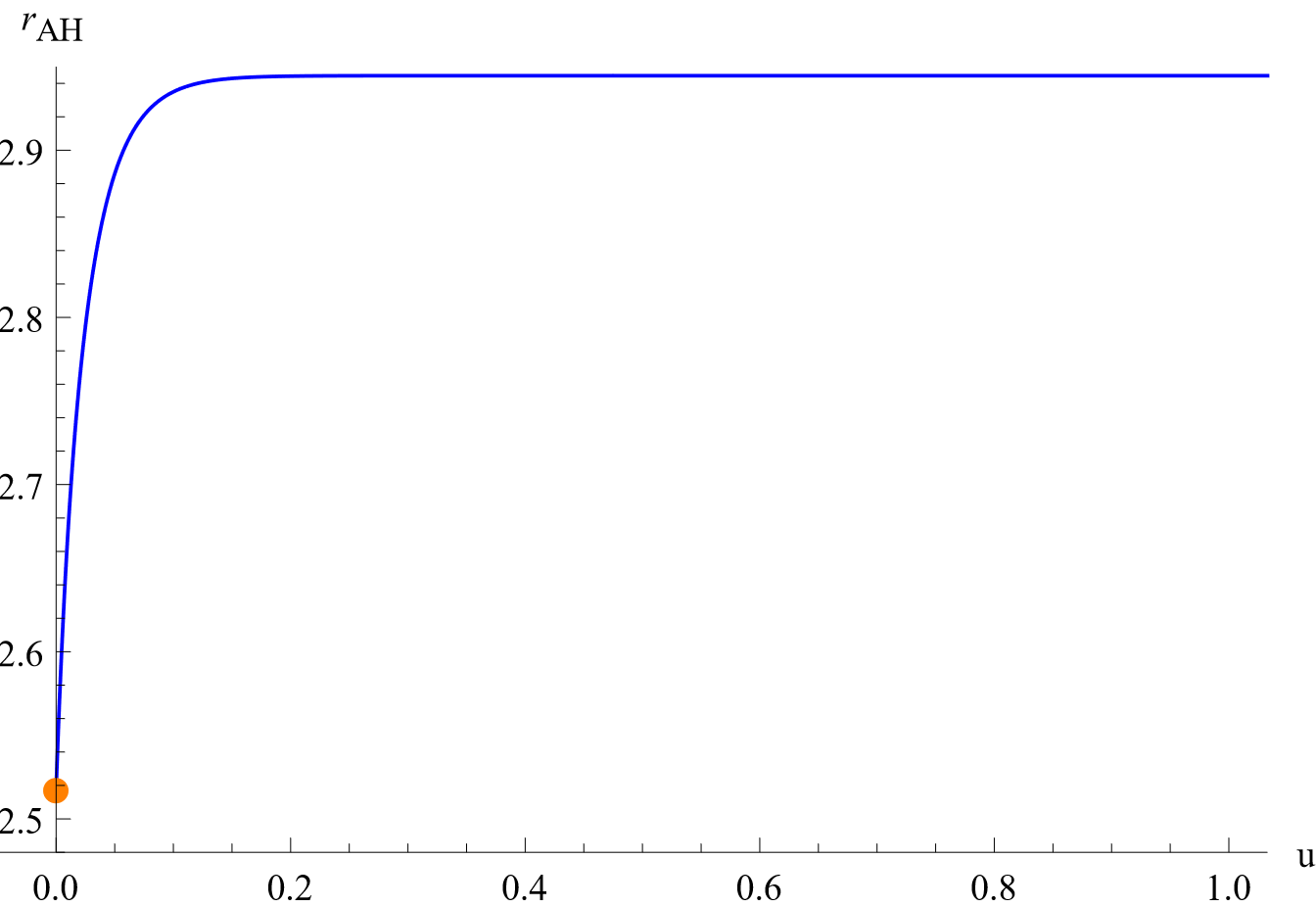}
  \caption{Left panel: The evolution of the scalar function $a(u)$ for the four dimensional solution with $k_0=1$ and $\gamma_2<0$. We have set $\gamma_2=-1\,,\alpha=5\,,c=\ft{7}{\sqrt{15}}\,,Q=4\sqrt{11}$ so that $\hat{q}_-=1\,,q_+=2$. The red point denotes the initial scalar charge $q_i=1.1>\hat{q}_-$. Right panel: The evolution of the apparent horizon. We further set $g=1\,,\gamma_1=-10/9$. The location for the event horizon for the initial black hole and the final stable black hole are
  $r_{h_1}\simeq 0.6711\,,r_{h_2}\simeq 2.9446$, respectively. However, the initial value of the apparent horizon (denoted by the orange point) is $r_{AH}(0)\simeq 2.5173\gg r_{h_1}$. This is a reminiscent of the fact $\dot{a}(0)>0$.}
  \label{d42}
\end{figure}

It is interesting to note that now the initial time of the evolution must be finite. For convenience, we choose the constant $u_0$ properly so that the initial time is zero. We have
\bea\label{d4jie22}
\ft34 \alpha\big(q_+^2+\tilde{q}_-^2 \big)u&=&q_+\, \Big[\mathrm{arctanh}\big(\fft{a}{q_+} \big)-\mathrm{arctanh}\big(\fft{q_i}{q_+} \big)\Big]\nn\\
&&+\tilde{q}_-\, \Big[\mathrm{arctan}\big(\fft{\tilde{q}_-}{a} \big)-\mathrm{arctan}\big(\fft{\tilde{q}_-}{q_i} \big)\Big]\,.
\eea
Here one may worry about what is the precise value of the initial scalar charge $q_i$. In fact, the positivity of the Vaidya mass constrains it to be
$\hat{q}_-<q_i<\hat{q}_+$ according to Eq.(\ref{d4scalar}). In addition, it is easy to show that for any electric charge in the range $0<Q^2\leq |9\alpha c^4/\gamma_2|$, one always has $\hat{q}_- \leq q_+$ where the equality is taken when the upper bound on the charge $Q$ is saturated. Thus, $q_i$ should be taken in the regiem $(\hat{q}_-\,,q_+)$. The dynamical evolution of the scalar function $a(u)$ as well as the apparent horizon is shown in Fig.\ref{d42}.
It follows that for any given initial scalar charge in this regime, the state is unstable because of $\dot a>0$. As a consequence, the initial Vaidya mass suddenly increases by an amount of $\delta M=q_i^2 \dot a/4\pi>0$ so that the apparent horizon is larger than the event horizon of the initial black hole.

\section{Conclusion}

In this paper, we constructed a class of exact static/dynamic black hole solutions in EMD theories. In the theory we have considered, the scalar field is nonminimally coupled to the Maxwell field with a generic function $Z(\phi)$. In addition, we further considered a general scalar potential $V(\phi)$.
We constructed exact black hole solutions using the reverse engineering procedure by proposing that the scalar $\phi$ took the form
$\phi=2k_{0}\,\text{arcsinh}\big[(\frac{q}{r})^{\Delta}\big]$. With this assumption, we can determine the black hole metric functions as well as the structure of the $Z(\phi)$ and $V(\phi)$.  The black hole solutions all involve the mass and the electric charge, and they are integration constants of the solutions rather than being specified by the parameters of the theories.

The integration constant $q$ can also be viewed as the scalar charge, and it is not a conserved quantity.  We therefore promoted $q$ to be a time dependent function $a(u)$ in the Eddington-Finkelstein-like coordinates and obtained a class of exact dynamic solutions where the static solution is the end point of the time evolution. The evolution is controlled by the evolution equation of $a$. We offered two examples to show that our solutions describe an initial black hole against nonlinear perturbations evolving into a final stable black hole.

\section*{Acknowledgement}

Z.Y.~Fan is supported in part by the National Natural Science Foundations of China (NSFC) with Grant No. 11805041, No. 11873025 and No. 11575270. H.H.~and H.L.~are supported in part by NSFC Grants No.~11875200 and No.~11475024. H.H.~is grateful to the Center for Joint Quantum Studies for hospitality.

\end{document}